# Open data, open review and open dialogue in making social sciences plausible

Quan-Hoang Vuong

Email: qvuong@ulb.ac.be

Centre Emile Bernheim, Universite Libre de Bruxelles (Belgium), and
Centre for Interdisciplinary Social Research, Western University Hanoi (Vietnam)

* **Abstract:**

Nowadays, protecting trust in social sciences also means engaging in open community dialogue, which helps to safeguard robustness and improve efficiency of research methods. The combination of open data, open review and open dialogue may sound simple but implementation in the real world will not be straightforward. However, in view of Begley and Ellis's (2012) statement that, "the scientific process demands the highest standards of quality, ethics and rigour," they are worth implementing. More importantly, they are feasible to work on and likely will help to restore plausibility to social sciences research. Therefore, I feel it likely that the triplet of open data, open review and open dialogue will gradually emerge to become policy requirements regardless of the research funding source.

* **Keywords:** Science policy; Data sharing; Society dialogs

* **JEL Code:** O38

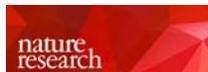
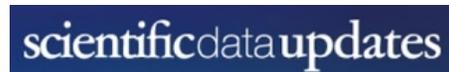

# Open data, open review and open dialogue in making social sciences plausible

Quan-Hoang Vuong

A growing awareness of the lack of reproducibility has undermined society's trust and esteem in social sciences. In some cases, well-known results have been fabricated or the underlying data have turned out to have weak technical foundations.

Many researchers have investigated the plausibility of findings in the social sciences and humanities. A typical example is the mysterious Critical Minimum Positivity Ratio 2.9013 by Fredrickson and Losada (2005), which claimed to show that there exists such a positivity ratio and that "an individual's degree of flourishing could be predicted by that person's ratio of positive to negative emotions over time". This ratio had once been a well-known, highly influential and greatly admired psychological "constant" until it was shown by Brown, Sokal and Friedman (2013) to be an unfounded, arbitrary and meaningless number.

To address the plausibility problem, I suggest that a combination of open data, open peer-review and open community dialogue, could serve as a feasible option for the social sciences.

**Open data**

Ethical standards for open data can be difficult to meet. As I discuss below, my own experience of publishing with *Scientific Data* (Vuong 2017) showed me firsthand the issues with managing ethical concerns. However, addressing these issues is not impossible, and underneath the difficulty and challenges lies the beauty of shared values and open thinking.

Still, psychological and financial barriers to open data remain. Although it sounds simple, the practice of pursuing an open data policy has turned out to be complicated and hard to achieve, so that "nudging scientific practices toward greater openness requires complementary and coordinated efforts from all stakeholders" (Nosek *et al* 2015). Unfortunately, the scientific community has not always taken this seriously (Begley and Ellis 2012), forsaking the opportunity for their valuable data to stand the test of time.

Now with the existence of reliable data repositories such as Harvard Dataverse, Open Science Framework, Mendeley, UK Data Archive, depositing data for public use and replication has never been easier. Open publishing platforms such as *F1000Research* even helps to securely deposit multiple data sets to their own systems without reliance on a third party's service.

*Scientific Data* goes even further to the ideal of removing restrictions on use of open data for commercial purposes, by advocating the practice of generating economic value through commercial enterprise, and showcasing the potential benefit for society and the economy (*Scientific Data* Editorial, "Open for business").

A benefit to authors of sharing data with a scholarly article is the opportunity for their data to be examined by other scholars, and possibly even lead to further scientific discoveries. The authors have the opportunity of having their colleagues verify both the soundness of data and robustness of the scientific process—as part of a cumulative process of uncertainty reduction (Aarts *et al* 2015). This is how the scientific workhorse of self-correction operates.

**My own experiences of open data**

Sometimes the process of data reuse can be quite fast, as is my own experience. When publishing a data article in *Data in Brief*, with open data sets deposited to Mendeley Data (Vuong and Nguyen, 2016), I was not sure if the data would be reused. To my surprise, a researcher at the University of Pittsburgh used them to make a quality contribution to *International Journal for Equity in Health* (Nguyen 2017). How quickly did this all happen? My paper was published on 30 September 2016 and Nguyen's manuscript was submitted on 15 November 2016!

Publishing data that was later reused was a special experience. Work value is transmitted via rigorous peer-review, to data users, i.e., other researchers, for the sake of advancing human knowledge. My own questions about the data's value were answered not just once, with the editorial and peer-review process, but twice and perhaps more than twice, with subsequent scientific publications that used the contributed data.

My subsequent experience while preparing data for publication at *Scientific Data* was even more special and unique, in the sense that meeting the most rigorous and demanding criteria boosted the sharpness, clarity, transparency, ethical values and reusability of my data. Even a highly accomplished author can find that the publishing process at *Scientific Data* enhances their own learning attitude and scientific exploration.

**Open review**

In the course of encouraging transparency and open science for the sake of better science for all, **open review** has emerged to be a commendable solution. This may sound strange to researchers in social sciences, where double-blind review has been predominant; however, theoretically the system has its advantages. Evidence from Walsh, Rooney, Appleby and Wilkinson (2000) supports open peer-review since signed reviews are of higher quality, more courteous, and more thoughtful. In addition, social values of open peer-review cannot be ignored as "Improvement of journal policies can help those values become more evident in daily practice and ultimately improve the public trust in science, and science itself," (Nosek *et al* 2015).

Heavyweight funders also see the value in the transparent peer-review process, public judgments, and the system's ability to address the issues of interdisciplinary research review as raised in Bammer (2016). That is why the young platform *F1000Research* received substantial support not long after its debut (Van Noorden, 2013), from Wellcome Trust in 2016 (Butler 2016) and Bill & Melinda Gates Foundation in 2017 (Butler 2017). As Begley and Ellis (2012) put it, social sciences now need that open review system, which helps facilitate a transparent discovery process leading to significant social benefits.

**Open community dialogue**.

We refer to an enlarged notion of "dialogue" that consists of technical expert discussions about scientific methods and computer codes (Eglen *et al* 2017) and the research communication processes that can be made available to the community for evaluation, critique, reuse or extension (Nosek *et al* 2015).

PubPeer has further incentivized the need for open community dialogue, as flagging a paper on the site is now perceived (incorrectly) by many as a threat. This is a technical expertise community that can help fix problems, deal with statistical weaknesses (that is to improve overall quality of manuscripts), and serve the quality gatekeeping for scientific outlets. PubPeer has been pushing what Eglen *et al* (2017) advocate, "Share the methods and computer codes." In social sciences, their values help update and verify, "stylized facts," in a Bayesian probabilistic world where, "an erroneous argument does not necessarily lead to a wrong conclusion," due to Gödel's theorem (Jaynes 2003).

Nowadays, protecting trust in social sciences also means engaging in open community dialogue, which helps to safeguard robustness and improve efficiency of research methods. The combination of open data, open review and open dialogue may sound simple but implementation in the real world will not be straightforward. However, in view of Begley and Ellis's (2012) statement that, "the scientific process demands the highest standards of quality, ethics and rigour," they are worth implementing. More importantly, they are feasible to work on and likely will help to restore plausibility to social sciences research. Therefore, I feel it likely that the triplet of open data, open review and open dialogue will gradually emerge to become policy requirements regardless of the research funding source.

**References**

Aarts AA *et al* (2015) Estimating the reproducibility of psychological science. *Science*; **349** (6251): aac4716, doi:10.1126/science.aac4716.

Bammer G (2016) What constitutes appropriate peer review for interdisciplinary research? *Palgrave Communications*; **2**: 16017, doi:10.1057/palcomms.2016.17.

Begley CG and Ellis LM (2012) Drug development: Raise standards for preclinical cancer research. *Nature*; **483** (7391): 531–533, doi:10.1038/483531a.

Brown NJL, Sokal AD, and Friedman HL (2013) The complex dynamics of wishful thinking: the critical positivity ratio. *American Psychologist*; **68** (9): 801–813, doi:10.1037/a0032850.

Butler D (2016) Wellcome Trust launches open-access publishing venture. *Nature News*

Butler D (2017) Gates Foundation announces open-access publishing venture. *Nature*; **543** (7647): 599, doi:10.1038/nature.2017.21700.

Eglen SJ *et al* (2017) Toward standard practices for sharing computer code and programs in neuroscience. *Nature Neuroscience*; **20** (6): 770-773, doi:10.1038/nn.4550.

Fredrickson BL and Losada MF (2005) Positive affect and the complex dynamics of human flourishing. *American Psychologist*; **60** (7): 678-686, doi:10.1037/0003-066X.60.7.678.

Jaynes ET (2003) *Probability Theory: The Logic of Science*. London, UK: Cambridge University Press.

Nguyen HM (2017) Inequality in healthcare costs between residing and non-residing patients: evidence from Vietnam. *Int. Journal for Equity in Health*; **16** (1): 76, doi:10.1186/s12939-017-0581-3.

Nosek BA *et al* (2015) Promoting an open research culture. *Science*; **348** (6242): 1422-1425, doi:10.1126/science.aab2374.

Scientific Data Editorial (2017) Open for business. *Scientific Data*; **4**: 170058, doi:10.1038/sdata.2017.58.

Van Noorden R (2013) Company offers portable peer review. *Nature*; **494** (7436): 161, doi:10.1038/494161a.

Vuong QH (2017) Survey data on Vietnamese propensity to attend periodic general health examinations. *Scientific Data*; **4**: 170142, doi:10.1038/sdata.2017.142.

Vuong QH and Nguyen TK (2016) Data on Vietnamese patients' financial burdens and risk of destitution. *Data in Brief*; **9**: 543-548, doi:10.1016/j.dib.2016.09.040.

Walsh E, Rooney M, Appleby L, and Wilkinson G (2000) Open peer review: a randomised controlled trial. *British Journal of Psychiatry*; **176** (1) 47-51; doi:10.1192/bjp.176.1.47.